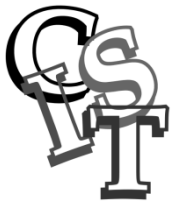

# معرفی یک الگوی جدید پیش‌توزیع کلید مبتنی بر طرح ترکیبیاتی PBIBD – μ در شبکه IoT با منابع محدود

اکبر مرشد اسکی[1]*، سید حمید حاج سید جوادی[2]

*نویسنده مسئول، دریافت: ۱۳۹۹/۰۹/۲۳، بازنگری: ۱۳۹۹/۰۹/۲۶، پذیرش: ۱۳۹۹/۱۰/۱۰

[1] استادیار، گروه کامپیوتر، دانشکده فنی و مهندسی، دانشگاه ورامین - پیشوا، دانشگاه آزاد اسلامی، ورامین، تهران، ایران

[2] استاد، دانشکده ریاضی و علوم کامپیوتر، دانشگاه شاهد، تهران، ایران

## چکیده

در شبکه IoT با منابع محدود از گره‌های انتهایی همچون حسگر بی‌سیم، RFID و سیستم‌های نهان استفاده می‌شود که دارای محدودیت حافظه، پردازش و انرژی می‌باشند. یکی از راه‌حل‌های توزیع کلید در این نوع شبکه‌ها، استفاده از روش پیش توزیع کلید است که عمل توزیع کلید را به‌صورت آفلاین قبل از استقرار دستگاه‌های منبع محدود در محیط انجام می‌دهد. همچنین به‌منظور کاهش سربار محاسباتی و ارتباطی کشف کلید مشترک، استفاده از طرح‌های ترکیبیاتی در پیش توزیع کلید به‌عنوان یک راهکار در سال‌های اخیر پیشنهاد شده است. در این مطالعه، به معرفی و ساخت یک طرح ترکیبیاتی از نوع PBIBD – μ پرداخته می‌شود و چگونگی نگاشت این طرح را به‌عنوان الگوی پیش توزیع کلید در شبکه IoT با منابع محدود بیان می‌گردد. با استفاده از این الگوی پیش توزیع کلید، کلیدهای بیشتری برای ارتباط بین دو دستگاه در شبکه IoT حاصل می‌شود. بدین معنی که بین هر دو دستگاه در شبکه حداکثر q+2 کلید موجود خواهد بود که در آن q توانی از یک عدد اول است. دو دستگاهی می‌توانند ارتباط امن مستقیم داشته باشند که به‌جای یک کلید مشترک، q+2 کلید مشترک در زنجیره کلیدیشان داشته باشند که به‌این‌ترتیب باعث افزایش مقاومت الگوی پیش توزیع کلید در آن‌ها نسبت به روش‌های SBIBD، TD، Trade-KP، *UKP، *RD و PBIBD – μ D-2 خواهیم شد.

**کلمات کلیدی:** شبکه IoT با منابع محدود، طرح‌های ترکیبیاتی، PBIBD – μ، مقاومت

## ۱- مقدمه

مفهوم IoT نخستین بار توسط کوین اشتن در سال ۱۹۹۹ بیان گردید که در آن به اشیا هویت دیجیتال به‌منظور پردازش و تبادل اطلاعات داده شد. شبکه IoT در کاربردهای زیادی از جمله شهر و خانه هوشمند، سلامت و صنعت بکار گرفته شده است [۱].

همچنین در شبکه IoT با منابع محدود از گره‌های بسیار زیادی همچون حسگر بی‌سیم، RFID و سیستم‌های نهان استفاده شده است که این گره‌ها دارای محدودیت‌هایی از قبیل حافظه، توان محاسباتی و ارتباطی و انرژی می‌باشند. از ویژگی‌های دیگر این نوع شبکه، وجود تعداد زیادی از گره‌ها در بسیاری از کاربردها است.

توزیع کلید به‌عنوان بخشی مهمی از مدیریت کلید در شبکه IoT است. به عمل توزیع کلید بین دستگاه‌ها قبل از استقرار آنها، پیش توزیع کلید گفته می‌شود و ازآنجایی‌که این عمل به شکل آفلاین صورت می‌گیرد، لذا در مصرف انرژی گره‌ها صرفه‌جویی می‌گردد.

روش‌های پیش توزیع کلید به سه دسته اصلی تصادفی، قطعی و ترکیبی تقسیم می‌شوند [۲]. روش‌های تصادفی پیش توزیع کلید ابتدا توسط اسچنور و گلیگور در [۳] مطرح گردیده است. در روش‌های تصادفی، حلقه‌های کلید دستگاه‌ها به‌صورت تصادفی از یک مخزن کلید گرفته می‌شود. این عمل توسط یک مرکز توزیع کلید موسوم به KDC انجام می‌شود (فاز پیش توزیع کلید)، پس از این مرحله، دستگاه‌ها با پخش همگانی مشخصه حلقه کلیدهای خود به دستگاه‌های همسایه و بازگشت نتیجه جستجو می‌توانند کلید مشترک بین هر دو دستگاه همسایه را در صورت

وجود پیدا کنند (فاز کشف کلید مشترک) که در صورت وجود این کلید مشترک، دو دستگاه همسایه با استفاده از آن کلید مشترک، ارتباط امن و مستقیم برقرار می‌کنند. همچنین در صورت عدم وجود کلید مشترک بین حلقه کلیدهای دو دستگاه همسایه، این دو دستگاه بایستی از یک مسیر چند پرشی و با استفاده از دستگاه‌های میانی که با گره‌های همسایه کلید مشترک دارند، ارتباط امن غیرمستقیم داشته باشند (فاز کشف مسیر کلید). این روش پس از آن به الگوی EG معروف شد.

روش تصادفی دیگری برای پیش توزیع کلید در [۴]، توسط چان و دیگران موسوم به q - ترکیبی ارائه شد که مقاومت آن نسبت به روش EG بهبود بهتری پیدا کرد. بدین صورت که از یک مخزن کلید کوچک‌تر نسبت به الگوی EG در آن استفاده شد که باعث شد دستگاه‌ها، بیش از یک کلید مشترک مابین خود داشته باشند. دو دستگاهی که q کلید مشترک در حلقه کلیدهای خود داشتند قادر به ارتباط امن و مستقیم بودند. به‌علاوه آنها نشان دادند که اگر تعداد گره‌های تسخیر شده توسط مهاجم کم باشد مقاومت الگوی q - ترکیبی، بیشتر از طرح EG است. همچنین در [۵]، کیون با استفاده از تابع هش یک الگویی برای KPS ارائه دادند که با استفاده از آن مقاومت بهبود یافت اما در آن تضمینی ارائه نشد که دو دستگاه کلید مشترکی داشته باشند، و لذا بایستی از مسیر چند پرشی برای ارتباط استفاده می‌کردند. در [۶]، لی و دیگران یک آستانه‌ای برای پیش توزیع کلید تصادفی ارائه نمودند که هر دستگاه در آن شبکه، با یک مسیر کلید امن به طول l می‌توانست با دستگاه همسایه دیگر ارتباط برقرار کند. همچنین در [۷]، با استفاده از چند مخزن کلید مجزا، میزان مقاومت در شبکه افزایش دادند.

الگوهای قطعی پیش توزیع کلید به سه دسته ماتریسی و چندجمله‌ای و ترکیبیاتی تقسیم می‌شود. الگوهای ماتریسی و چند جمله‌ای در مقالات [۸-۱۱] مورد بررسی قرار گرفته‌اند اما مشکل عمده در آنها این است که دارای امنیت آستانه‌ای بودند. بدین معنی که با تسخیر t دستگاه تمام ارتباطات در شبکه ناامن می‌شود.

در پیش‌توزیع کلید با استفاده از طرح‌های ترکیبیاتی، به علت استفاده از یک الگوریتم کاملاً مشخص در تولید حلقه‌های کلید، سربار محاسباتی و ارتباطی کشف کلید مشترک کاهش می‌یابد. همچنین میزان اتصال‌پذیری را می‌توان با انتخاب طرح مناسب، تغییر داده و درعین‌حال مقاومت مناسبی را برای شبکه IoT فراهم نمود.

اولین الگوی پیشنهادی مبتنی بر طرح‌های ترکیبیاتی در [۱۲] مطرح شد که در آن از طرحی ترکیبیاتی موسوم به طرح متقارن استفاده شد. الگوی پیش توزیع کلید پیشنهادی دارای اتصال‌پذیری کامل بوده، اما از لحاظ مقاومت و مقیاس‌پذیری چندان مطلوب نبوده است. در [۱۳]، لی و دیگران با استفاده از طرح ترکیبیاتی TD الگوی پیش توزیع کلیدی معرفی کردند که مقاومت بهتری را نسبت به طرح متقارن ایجاد می‌کرد. در [۱۴] و [۱۵] طرح‌های ترکیبیاتی دیگری به‌منظور افزایش مقاومت به کار گرفته شده است. در [۱۶]، بچکیت و دیگران از طرح ترکیبیاتی دیگری موسوم به طرح Unital، به‌منظور افزایش مقیاس‌پذیری و اتصال‌پذیری بهتر استفاده کردند. همچنین در [۱۷]، مدیری و دیگران با استفاده از طرح Residual (RD) یک الگوی پیش توزیع کلیدی را ارائه نمودند که در آن میزان مقیاس‌پذیری و اتصال‌پذیری نسبت به طرح‌های Trade و Unital بهبود داده شد. همچنین در [۱۸]، یان و دیگران، طرحی مبتنی بر طرح ترکیبیاتی μ-PBIBD روی فضای دو بعدی و سه بعدی ارائه داده‌اند و نشان دادند در حالت سه بعدی، این طرح نسبت به طرح ارائه شده در [۱۷] امنیت، بیشتر و اتصال‌پذیری بالاتری دارد علاوه بر آنکه از مقیاس‌پذیری بالایی نیز برخوردار است.

الگوهای پیش توزیع کلید ترکیبی از ترکیب الگوهای قطعی و تصادفی حاصل شده است. این الگوها علاوه بر سودمندی الگوهای تصادفی از لحاظ مقاومت و مقیاس‌پذیری، دارای سودمندی الگوهای قطعی که بیشتر به‌منظور بهبود اتصال‌پذیری و سربار کمتر کشف کلید مشترک بین دو دستگاه IoT می‌باشد، را دارا

هستند. در [۱۹] و [۲۰] الگوی ترکیبی مبتنی بر طرح‌های ماتریسی و چندجمله‌ای با طرح تصادفی EG ارائه شد. همچنین در [۲۱-۲۳] و [۲۰] از الگوهای پیش توزیع کلید ترکیبی مبتنی بر طرح‌های ترکیبیاتی به‌منظور دستیابی به اتصال‌پذیری و مقاومت بهتر استفاده کردند.

در این کار یک طرح ترکیبیاتی جدیدی از نوع μ-PBIBD روی طرح متقارن تعریف کرده و با نگاشت آن به پیش توزیع کلید در شبکه IoT با منابع محدود به الگوی پیش توزیع کلیدی دست می‌یابیم که بین هر دو گره آن بیش از یک کلید برای ارتباط وجود دارد. دو دستگاه همسایه در آن می‌توانند دارای ارتباط مستقیمی باشند که q+2 کلید مشترک بین حلقه کلید خود داشته باشند. بدین ترتیب به الگوی پیش توزیع کلیدی با مقاومت و مقیاس‌پذیری بالایی دست می‌یابیم.

## ۱-۲- طرح‌های ترکیبیاتی

اگر X مجموعه‌ای از v عنصر متمایز و A یک مجموعه متناهی d عضوی موسوم به بلوک، از زیرمجموعه‌های X باشد. به زوج (X,A) یک طرح ترکیبیاتی یا طرح بلوکی گویند. به تعداد بلوک‌های شامل عضو $x \in X$، درجه عضو x هستند. اگر تمام اعضای X از درجه r باشند، طرح (X,A)، منظم نامیده می‌شود. به‌اندازه بزرگ‌ترین بلوک در طرح بلوکی مرتبه گفته می‌شود و درصورتی‌که تمام بلوک‌ها دارای تعداد اعضای برابر، k، باشند، آنگاه، این طرح را یکنواخت از مرتبه k نامیده می‌شود. طرحی که در آن r = k = d و v = d باشد یک طرح متقارن نامیده می‌شود[۲۴].

مثال۱: زوج (X , A) که در آن
X = {1, 2, 3, 4, 5, 6, 7, 8, 9}
A={{1,2,3}, {4,5,6}, {7,8,9}, {1,4,7}, {2,5,8}, {3,6,9}, {1,5,9}, {2,6,7}, {3,4,8}, {1,6,8}, {2,4,9}, {3,5,7}}
طرحی است که از ۹ نقطه و ۱۲ بلوک تشکیل شده است. این طرح، منظم از درجه ۴ و یکنواخت از مرتبه ۳ است.

تعریف۱: به طرح بلوکی (X,A) که در آن هر زوج از عناصر X دقیقاً در λ بلوک از مجموعه A موجود باشند (شرط بالانس بودن)، به آن یک BIBD می‌نامند و با نماد B(v,d,r,k,λ) نشان می‌دهند. v و r و d و k پارامترهای BIBD هستند که در رابطه dk=vr و λ(v-1)=r(k-1) صدق می‌کنند. به‌ویژه اگر d=v باشد، آنگاه r=k خواهد بود. در این صورت به BIBD حاصل، BIBD متقارن (SBIBD) گفته می‌شود و با نماد SBIBD(v,r,λ) نشان داده می‌شود.

قضیه۱: برای هر توانی از یک عدد اول مانند q که در آن $q \geq 2$ باشد یک SBIBD با پارامترهای SBIBD($q^2$ +q+1,q,1) وجود دارد. طریقه ساخت این SBIBD در [۲۱] و [۲۵] آورده شده است.

مثال۲: اگر V={1,2,3,4,5,6,7} و B = {B_1, B_2, ... , B_v} که در آن $B_1 = \{1,2,3\}$، $B_2 = \{1,4,5\}$، $B_3 = \{1,6,7\}$، $B_4 = \{2,4,6\}$، $B_5 = \{2,5,7\}$، $B_6 = \{3,4,7\}$ و $B_7 = \{3,5,6\}$. در این صورت یک SBIBD به شکل SBIBD(7,3,1) با مقدار q=2 خواهیم داشت.

تعریف۲: در طرح BIBD اگر هر زوج از عناصر X در تعداد متفاوتی از بلوک‌های مجموعه A موجود باشند (شرط بالانس جزئی بودن) به آن یک BIBD جزئی (PBIBD) می‌گویند. به‌عبارت‌دیگر مجموعه‌ای مانند $F = \{\lambda_1, \lambda_2, ..., \lambda_\mu\}$ از اعداد صحیح مثبت باشد که هر زوج از نقاط V در $\lambda_i$ بلوک برای هر $1 \leq i \leq \mu$ ظاهر شود. در این صورت μ-PBIBD با پارامترهای μ-PBIBD(v,d,r,k,λ_1,λ_2,...λ_μ) نشان داده می‌شود.

قضیه ۲: اگر $\lambda_1=\lambda_2=...=\lambda_\mu$ باشد μ-PBIBD به یک BIBD تبدیل می‌شود.

مثال۳: روی مجموعه X={1,2,3,4,5,6} با درنظرگرفتن مجموعه بلوک‌های A={{1,4,2,5},{2,5,3,6},{3,6,1,4}}، یک μ-PBIBD (6,3,2,4,1,2) به دست می‌آید که در آن μ=2 است. بدین صورت که این طرح ۳ بلوکی دارا

روی مجموعه ۶ عضوی X دارد و هر عضو از X در دو بلوک و هر بلوک ۴ عضو و همچنین هر دو عضو X در ۱ یا ۲ بلوک قرار دارد.

یک طرح Transversal با نماد TD(k,q) نشان داده می‌شود. برای این‌که TD(k,q) قابل ساختن باشد، q باید توانی از یک عدد اول و $2 \leq k \leq q$ در نظر گرفته شود. ابتدا یک مجموعه kq عضوی بنام V، در نظر گرفته می‌شود. سپس دو مجموعه بنام‌های، مجموعه گروه‌ها و مجموعه بلوک‌ها بدین صورت ساخته می شود که مجموعه گروه‌ها یک مجموعه k عضوی از q- زیرمجموعه‌های مجزای V، است که V را افراز می‌کند و مجموعه بلوک‌ها، یک مجموعه از k-زیرمجموعه‌های مجموعه V در نظر گرفته می‌شود به طوری که

۱- هر بلوک دقیقاً شامل یک عضو از هر گروه باشد.

۲- هر زوج عنصر از دو گروه مختلف دقیقاً در یک بلوک قرار داشته باشند.

TD(k,q) دارای خواص زیر است:

۱- در این طرح تعداد kq نقطه داریم که در $q^2$ بلوک قرار گرفته‌اند.

۲- هر بلوک دقیقاً k نقطه دارد.

۳- هر نقطه دقیقاً در q بلوک قرار دارد.

مثال۴: اگر X={1,2,3,...,12} باشد یک TD(4,3) دارای گروه‌هایی مانند زیر است:

G$_1$={1,2,3} , G$_2$={4,5,6} , G$_3$={7,8,9} , G$_4$={10,11,12}

و بلوک‌های آن به‌صورت زیر می‌باشند.

{1,4,7,10}, {1,5,8,11}, {1,6,9,12}, {2,4,8,12}, {2,5,9,10},
{2,6,7,11} , {3,4,9,11} , {3,5,7,12} , {3,6,8,10}

در [۲۶]، ساخت طرح TD بیان شده است.

### ۱-۳- ساختار مقاله

سازماندهی مقاله در ادامه بدین صورت است که، ابتدا در بخش ۲ به کارهای مرتبط پرداخته و بعد از آن، در بخش ۳ مدل پیشنهادی مورد نظرمان را معرفی می‌نماییم. سپس در بخش‌های ۴ و ۵ به ترتیب به تحلیل و مقایسه کارایی مدل پیشنهادی پرداخته و در نهایت در بخش ۶ نتیجه‌گیری آورده می‌شود.

## ۲- کارهای مرتبط

چندین الگوی پیش‌توزیع کلید مبتنی بر طرح‌های ترکیباتی وجود دارد. اولین آنها از این نوع، توسط کامتپ و ینر در [۲۱]، ارائه گردیده است که برای پیش‌توزیع کلید از یک SBIBD با پارامترهای (SBIBD(q$^2$ +q+1,q,1 که در آن q توانی از یک عدد اول است، استفاده کردند و $q^2 + q + 1$ بلوک این طرح را به‌عنوان زنجیره کلید به دستگاه‌ها نسبت دادند. در این الگوی پیش‌توزیع کلید با طول زنجیره کلید k=q+1، تعداد $q^2 + q + 1$ دستگاه IoT قابل پوشش است. همچنین این الگو دارای خاصیت اتصال‌پذیری کامل بود. بدین معنی که هر دو دستگاه در شبکه، دارای یک کلید مشترک برای ارتباط بودند، اما مقیاس‌پذیری و مقاومت پایینی را داشت.

در [۱۵]، Ruj و دیگران یک الگوی پیش‌توزیع کلید موسوم به Trade-KP ارائه نمودند. یک Trade (v,k)-t موسوم bitrade از دو گردایه m عضوی T$_1$ و T$_2$ از زیرمجموعه‌های k عضوی از X با v عنصر تشکیل شده‌اند، به طوری که $T_1 \cap T_2 = \emptyset$. و هر زیرمجموعه t عضوی از X دقیقاً در تعداد یکسانی از بلوک‌های T$_1$ و T$_2$ قرار دارد. m به‌عنوان اندازه Trade نامیده می‌شود.

یک Trade (v,k)-t را از نوع Steiner گویند، هرگاه هیچ زیرمجموعه t عضوی بیش از یکبار در T1 یا T2 ظاهر نشود. و Trade (v,k)-2 از نوع Strong Steiner Trade یا به‌اختصار SST گویند هرگاه هر بلوک از T$_1$ با هر بلوک از T$_2$ حداکثر در ۲ عضو مشترک باشند. به‌عنوان‌مثال با فرض اینکه X={1,2,3,4,5,6} باشد، T={T$_1$,T$_2$} یک Trade (6,3)-2 از نوع SST است.که در آن

T1 = {{1, 2, 3}, {1, 5, 6}, {2, 4, 6}, {3, 4, 5}}
T2 = {{1, 2, 6}, {1, 3, 5}, {2, 3, 4}, {4, 5, 6}}

در [۲۳] با توجه به وجود یک SST از نوع (1 + q ,1 + q + $q^2$) - 2 درصورتی‌که q توانی از یک عدد اول باشد. بلوک‌های $T_1 \cup T_2$ را به‌عنوان زنجیر کلید به دستگاه‌ها اختصاص می‌دهند. و تعداد دستگاه‌های قابل پوشش در این طرح (1 + q + $q^2$)2 است و هر دستگاه دارای طول زنجیره کلید q+1 است. همچنین آنها نشان دادند که، مقاومت طرح ارائه شده از الگوهای قطعی قبل مانند SBIBD و طرح TD در [۲۶] و PBIBD در [۱۴] بیشتر است. اما فاقد خاصیت اتصال‌پذیری کامل است.

بچکیت و دیگران در [۱۶]، یک الگوی پیش‌توزیع کلید مبتنی بر طرح Unital ارائه دادند و آن را به‌اختصار NU-KP نامیدند. طرح Unital یک BIBD، با پارامترهای (1,1 + q ,$q^2$ ,(1 + q - $q^2$)$q^2$ ,1 + $q^3$)BIBD است. همچنین به‌منظور افزایش احتمال کلید مشترک و مقاومت طرح آنها یک الگوی پیش‌توزیع کلید موسوم به t-UKP معرفی کردند، که در آن به هر دستگاه، t بلوک متمایز از طرح NU-KP را به‌عنوان زنجیره کلید نسبت دادند و با انتخاب $t = \sqrt{q}$، طرح را *UKP نامیدند. با توجه به نتایج ارائه در آن مقاله، هرچند طرح‌های ارائه شده مقیاس‌پذیری بهتری نسبت به SBIBD داشتند، اما از مقاومت کمتری نسبت به آن برخوردار بودند.

مدیری و دیگران در [۱۷]، برای ساخت طرح Residual، از یک SBIBD با پارامترهای (1,1 + q + $q^2$)SBIBD استفاده کردند. بدین صورت که با فرض اینکه بلوک‌های طرح SBIBD ساخته شده $B = \{B_1, B_2, B_3, ... B_b\}$ باشد که در آن تعداد بلوک‌ها $b = q^2 + q + 1$ است. در طرح Residual تعداد $b = (q^2 + q + 1)(q^2 + q)$ بلوک به‌صورت B$_{ij}$=B$_i$-B$_j$ که در آن i,j=1,2,3,...,b است، را می‌سازند. که به‌اختصار RD نامگذاری کردند و سپس در این مقاله به‌منظور افزایش مقاومت، بلوک‌های تکراری ایجاد شده در این طرح را حذف نمودند و آنرا به‌اختصار *RD نامیدند. و نشان دادند این طرح مقاومت و مقیاس‌پذیری بهتری نسبت به طرح‌های SBIBD و Trade-KP و *UKP دارد. با این حال این طرح نیز فاقد ویژگی اتصال‌پذیری کامل است.

Yuan و دیگران در [۱۸]، با ساخت یک PBIBD − μ با پارامترهای μ:

μ − PBIBD($q^2, q^2, 2q − 2, 2q − 2, \lambda_1 = 2, \lambda_2 = q − 2$)

که روی یک صفحه دو بعدی ساخته می‌شود طرحی موسوم به μ − 2D sPBIBD معرفی می‌کنند. که دارای خاصیت اتصال‌پذیری کامل بوده اما فاقد مقاومت مناسب نسبت به طرح‌های قبل بودند که با تعریف این طرح روی یک فضای سه بعدی و ساخت یک PBIBD − μ با پارامترهای:

μ − PBIBD($q^3, q^3, 3q − 3, 3q − 3, \lambda_1 = 0, \lambda_2 = 2, \lambda_3 = q − 2$)

طرحی موسوم به sPBIBD − μ 3D معرفی می‌کنند که مقاومت آن نسبت به طرح‌های SBIBD، طرح TD در [۲۶]، و طرح *RD در [۱۷] بیشتر بوده اما فاقد خاصیت اتصال‌پذیری کامل بوده است.

## ۳- مدل پیشنهادی

به‌منظور ساخت طرح ترکیباتی مبتنی بر μ-PBIBD و نگاشت آن به شبکه IoT با اندازه حافظه K کلید، ابتدا توانی از عدد اول q را طوری در نظر می‌گیریم که K≥(q+1)2. سپس با این q دو SBIBD1 و SBIBD2 با پارامترهای یکسان (1,1 + $q_1$ ,1 + $q_1^2$) روی دو مجموعه مجزا X$_1$ و X$_2$ می‌سازیم که هر یک از دو مجموعه به‌عنوان مخزن کلید مجزا محسوب می‌شود. حال یک مش دو بعدی با $(q_1^2 + q_1 + 1)^2$ سلول در نظر می‌گیریم که تعداد سطر و ستون آن برابر $q_1^2 + q_1 + 1$ در نظر گرفته شده است. بلوک‌های SBIBD1 را به‌عنوان بخش اول زنجیره کلید به سطرهای مش دو بعدی نسبت می‌دهیم و بلوک‌های SBIBD2 را به‌عنوان بخش دوم زنجیره کلید به ستون‌های مش دو بعدی نسبت می‌دهیم و در هر سلول اجتماع زنجیره کلید موجود در هر سر و ستون را قرار می‌دهیم. مشابه آنچه که در [۲۷] نشان داده شده است به سادگی می‌توان دید که طرح ترکیباتی حاصل یک μ-PBIBD با پارامترهای $\mu - \text{PBIBD}(v = 2(q^2 + q + 1), d = (q^2 + q + 1)^2, r = (q+1)(q^2 + q + 1), k = 2(q+1), \lambda_1 = (q^2 + q + 1), \lambda_2 = (q+1)^2)$ است. که بین هر دو بلوک در آن 2 یا q+2 کلید مشترک وجود دارد.

در این طرح ترکیباتی 2= μ است. همچنین این طرح بلوکی، $d = (q^2 + q + 1)^2$ بلوک روی مجموعه $v = 2(q^2 + q + 1)$ عضوی X دارد و هر عضو از X در $r = (q+1)(q^2 + q + 1)$ بلوک و هر بلوک $k = 2(q+1)$ عضو و هر دو عضو X در $\lambda_1 = (q^2 + q + 1)$ یا $\lambda_2 = (q+1)^2$ بلوک قرار دارد.

### ۳-۱- نگاشت طرح ترکیباتی مبتنی بر μ-PBIBD به دستگاه‌های شبکه IoT با منابع محدود

با فرض آنکه هر دستگاه IoT دارای حافظه‌ای با طول K کلید است، مش دو بعدی در مدل پیشنهادی را مطابق بخش قبل ساخته و در KDC قرار می‌دهیم.

در فاز توزیع کلید در مدل پیشنهادی برای هر دستگاه شبکه، زنجیره کلید موجود در یک سلول از مش دو بعدی توسط KDC به‌عنوان زنجیره کلید به هر دستگاه نسبت داده می‌شود که در این انتساب علاوه بر زنجیره کلید، KDC موظف است شناسه‌ای را برای هر کلید در نظر گرفته و به همراه هر کلید در زنجیره کلید دستگاه‌ها ذخیره و به هر دستگاه IoT نسبت دهد. در این صورت شبکه‌ای با اندازه $N = (q^2 + q + 1)^2$ پوشش داده خواهد شد. که بین هر دو دستگاه در آن 2 یا q+2 کلید مشترک وجود دارد. که این مرحله به شکل آفلاین انجام می‌شود.

در فاز کشف کلید مشترک هر دستگاه شناسه‌های کلید موجود در حلقه کلید خود را به‌منظور کشف کلید مشترک، بین دستگاه‌های همسایه پخش همگانی می‌کند و دو دستگاهی که دارای q+2 کلید مشترک باشند، می‌توانند ارتباط مستقیم و امن برقرار کنند. در غیر این صورت فاز کشف کلید مسیر را اجرا می‌نمایند. در حقیقت اگر حلقه کلیدهای توزیع شده بین دو دستگاه از یک سطر یا یک ستون یکسانی از مش دو بعدی باشد دو دستگاه q+2 کلید مشترک خواهند داشت.

در فاز کشف مسیر کلید درصورتی‌که نتوان بین دو گره همسایه برای ارتباط q+2 کلید مشترک پیدا نمود. دو دستگاه همسایه از طریق دستگاه‌های میانی که با آنها q+2 کلید مشترک دارند و از طریق یک مسیر چند پرشی ارتباط امن برقرار می‌کنند. در حقیقت اگر حلقه کلیدهای توزیع شده بین دو دستگاه از سطر و ستون متفاوتی از مش دو بعدی باشند، دو دستگاه تنها ۲ کلید مشترک خواهند داشت. اما دو دستگاه یافت می‌شود که دارای سطر یکسان با یکی و ستون یکسان با دستگاه دیگر می‌باشد و لذا با هر یک از دو دستگاه همسایه q+2 کلید مشترک دارند.

همچنین برای امنیت بیشتر به‌جای استفاده از q+2 کلید برای ارتباط، از hash آنها برای ارتباط استفاده می‌شود.

## ۴- تحلیل کارایی

### ۴-۱- احتمال داشتن q+2 کلید مشترک

$$pro = \frac{2\binom{q^2+q+1}{1}\binom{q^2+q+1}{2}}{\binom{(q^2+q+1)^2}{2}} = \frac{2}{q^2+q+1} \quad (1)$$

### ۴-۲- مقاومت

ازآنجایی‌که هر کلید در $(q+1)(q^2+q+1)$ قرار دارد. q+2 کلیدی که برای یک ارتباط استفاده می‌شود حداکثر در تعداد عناصر یک سطر یا یک ستون از مش دو بعدی، یعنی در $(q^2+q+1)$ دستگاه دیگر موجود است، لذا احتمال لو نرفتن این ارتباط در صورت تسخیر x دستگاه از شبکه برابر

$$\frac{\binom{N-(q^2+q+1)}{x}}{\binom{N-2}{x}} \quad (2)$$

که در آن $N = (q^2+q+1)^2$،

بنابراین احتمال لو رفتن یک ارتباط در شبکه، در صورت تسخیر x دستگاه برابر

$$Res(x) = 1 - \frac{\binom{N-(q^2+q+1)}{x}}{\binom{N-2}{x}} \quad (3)$$

که در آن $N = (q^2+q+1)^2$.

### ۴-۳- مقیاس‌پذیری

با توجه به مشخصات مدل پیشنهادی بیان شده در بخش ۳، لذا تعداد دستگاه قابل پوشش توسط آن برابر $N = (q^2+q+1)^2$ که طول حلقه کلید هر دستگاه در آن برابر $2(q+1)$ است.

## ۵- مقایسه کارایی

### ۵-۱- مقیاس‌پذیری

جدول ۱ مقیاس‌پذیری الگوهای پیش توزیع کلید مختلف را با توجه به‌اندازه حلقه کلید هر طرح بیان می‌کند.

جدول۱: پارامترهای طرح‌های مختلف پیش توزیع کلید

| طرح ترکیباتی | اندازه مخزن کلید | تعداد حلقه‌های کلید | اندازه حلقه کلید |
|---|---|---|---|
| SBIBD [۲۱] | $q^2+q+1$ | $q^2+q+1$ | $q+1$ |
| TD [۲۶] | $kq$ | $q^2$ | $k$ |
| Trade-KP [۱۶] | $q^2+q+1$ | $2q^2$ | $q$ |
| t-UKP [۱۶] | $q^3+1$ | $\frac{q^2(q^2-q+1)-(t-1)(q^2-1)(q+1)}{t}$ | $q+1$ |
| RD* [۱۷] | $q^2+q+1$ | $(q^2+q+1)(q+1)$ | $q$ |
| 2-D $\mu$-PBIBD [۱۸] | $q^2$ | $q^2$ | $2q-2$ |
| 3-D $\mu$-PBIBD [۱۸] | $q^3$ | $q^3$ | $3q-3$ |
| مدل پیشنهادی | $2(q^2+q+1)$ | $(q^2+q+1)^2$ | $2(q+1)$ |

نمودار زیر که بر اساس مقادیر جدول فوق به ازای طول حلقه کلید، رسم گردیده و نشان دهنده مقیاس‌پذیری بیشتر مدل پیشنهادی نسبت به تمام طرح‌های مورد مقایسه است.

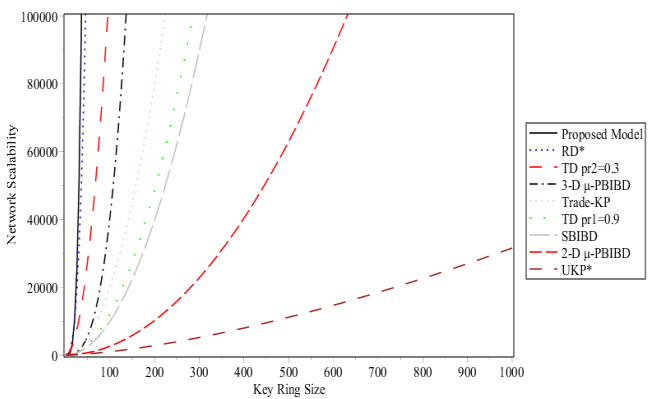

شکل۱: مقایسه مقیاس‌پذیری الگوهای مختلف پیش‌توزیع کلید

## ۵-۲- اتصال‌پذیری

در $SBIBD(q^2+q+1, q+1, 1)$، 2-D μ-PBIBD احتمال کلید مشترک بین هر دو دستگاه برابر ۱ است، یعنی اتصال‌پذیری کامل را داریم [۲۱،۱۸]. در طرح TD احتمال کلید مشترک برابر است با

$$\frac{k}{q+1} \quad (4)$$

که برحسب دو پارامتر k و q است [۲۶]؛ لذا به‌منظور مقایسه، اندازه شبکه را در این طرح با طرح پیشنهادی یکسان در نظر می‌گیریم.

در طرح Trade-KP، احتمال کلید مشترک برابر $\frac{q(q-1)}{2(2q^2-1)}$ است، و در طرح t-UKP، احتمال کلید مشترک برابر

$$1-(1-\frac{(q+1)}{(q^3+q+1)})^{t^2} \quad (5)$$

که در *UKP، مقدار t برابر $\sqrt{q}$ است [۱۶].

و در

$$RD^*(q^2+q+1, (q^2+q+1)(q+1), q(q+1), q, 1)$$

احتمال کلید مشترک طبق [۱۷]، از رابطه زیر به دست می‌آید.

$$P_{RD^*} = \frac{\binom{q^2+q}{2}}{\binom{(q^2+q)(q^2+q+1)}{2}} \times \left(\frac{q^2}{q^2+q}\right) + \left(\frac{(q^2+q)^2}{\binom{(q^2+q)(q^2+q+1)}{2}}\right) \times \left(\frac{q^4+q-1}{(q^2+q)^2}\right) \quad (6)$$

همچنین در 3-D μPBIBD مطابق [۱۸]، احتمال کلید مشترک برابر است با

$$\frac{3q}{q^2+q+1} \quad (7)$$

مقایسه نمودارهای احتمال کلید مشترک با زنجیره کلید یکسان، در شکل ۲ داده شده است؛ که بیانگر این است که مدل پیشنهادی اتصال‌پذیری بهتری نسبت به همه طرح‌های مقایسه شده به‌جز SBIBD و 2-D μ-PBIBD را دارد.

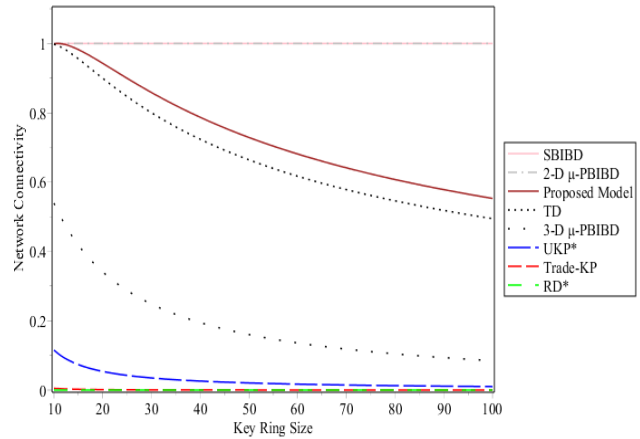

شکل۲: مقایسه اتصال‌پذیری الگوهای مختلف پیش‌توزیع کلید

## ۵-۳- مقاومت

طبق [۲۱]، [۲۶]،[۱۶]، [۱۷] و [۱۸]، مقاومت در طرح‌های SBIBD، TD، Trade-KP، t-UKP، $RD^*$، 2-D μ-PBIBD و 3-D μ-PBIBD طبق فرمول‌های زیر محاسبه می‌شود.

$$Res_{SBIBD}(x) = 1 - \frac{\binom{q^2}{x}}{\binom{q^2+q+1}{x}} \quad (8)$$

$$Res_{TD}(x) = 1 - \left(1-\frac{q-2}{q^2-2}\right)^x \quad (9)$$

$$Res_{Trade-KP}(x) = 1 - \frac{\binom{2q^2-4q+2}{x} + 4(q-1)\binom{2q^2-4q+2}{x-1}}{\binom{2q^2}{x}} \quad (10)$$

$$Res_{t-UKP}(x) = 1 - \sum_{i=1}^{t^2} \left(1 - \frac{\binom{q^3(q-1)}{xt}}{\binom{q^2(q^2-q+1)}{xt}}\right) \frac{\binom{t^2}{i}\left(\frac{(q+1)^2}{q^3+q+1}\right)^i \left(1-\frac{(q+1)^2}{q^3+q+1}\right)^{t^2-i}}{1-(1-\frac{(q+1)^2}{(q^3+q+1)})^{t^2}} \quad (11)$$

که در *UKP که $t=\sqrt{q}$ است.

$$Res_{RD^*}(x) = \sum_{j=1}^{q^2+q+1} \frac{\binom{q(q+1)}{2}}{\binom{(q^2+q+1)(q+1)}{2}} (1-\frac{\binom{(q+1)(q^2+q+1)}{x}}{\binom{(q^2+q+1)(q+1)}{x}}) \quad (12)$$

$$Res_{2-D\,\mu PBIBD}(x) = \frac{q-1}{q+1}\left(1 - 2\left(1-\frac{2q-4}{q^2-2}\right)^x + \left(1-\frac{4q-8}{q^2-2}\right)^x\right) + \frac{2}{q+1}(1 - \frac{\binom{q^2-q}{x} + \binom{q-2}{1}\binom{q^2-q}{x-1}}{\binom{q^2-2}{x}} + \frac{Ch(x)}{\binom{q(q-1)}{x}}) \quad (13)$$

که در آن

$$Ch(x) = \binom{q(q-1)}{x} - \binom{q-2}{1}\binom{(q-1)(q-1)}{x} + \binom{q-2}{2}\binom{(q-1)(q-2)}{x} + \cdots + (-1)^\theta \binom{q-2}{\theta}\binom{(q-1)(q-\theta)}{x} \quad (14)$$

که در آن $\theta \leq q-2$ و $q-2 \leq x \leq (q-1)(q-2)$

همچنین

$$Res_{3-D\ \mu PBIBD}(x) = \frac{3q-3}{q^2+q+1}\left(1 - 2\left(1 - \frac{3q-5}{q^3-2}\right)^x\right.$$
$$\left. + \left(1 - \frac{6q-10}{q^3-2}\right)^x\right) + \frac{3}{q^2+q+1}(1$$
$$- \frac{\binom{q^2-q}{x} + \binom{q-2}{1}\binom{q^2-q}{x-1}}{\binom{q^3-2}{x}} + \frac{Ch(x)}{\binom{q(q-1)}{x}})$$

(۱۵)

در شکل ۳ نمودار مقایسه مقاومت طرح‌های فوق با طرح پیشنهادی آورده شده است؛ که نشان‌دهنده این است که مقاومت طرح پیشنهادی نسبت به تمام طرح‌های مقایسه شده به جز 3-D μ-PBIBD بیشتر است.

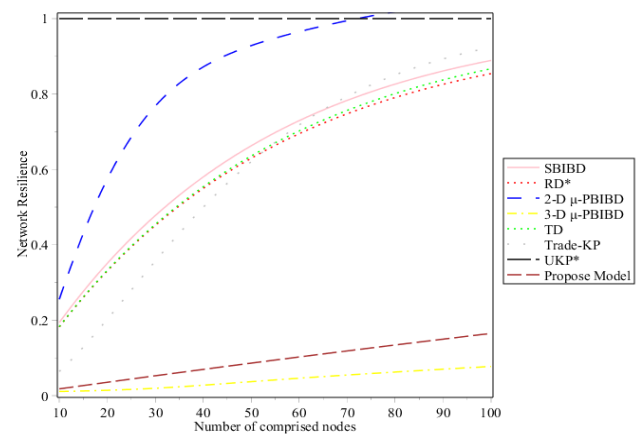

شکل۳: مقایسه مقاومت الگوهای مختلف پیش‌توزیع کلید

## ۶- نتیجه‌گیری

در این کار یک طرح ترکیبیاتی جدید از نوع $\mu - PBIBD$ معرفی شده و به‌عنوان الگوی پیش‌توزیع کلید به شبکه IoT با منابع محدود، نگاشت شده است. و به آنالیز طرح پیشنهادی به‌منظور بدست آوردن کارایی طرح پرداخته و نشان داده شد که با کمک این طرح ترکیبیاتی، به الگوی پیش‌توزیع کلیدی با مقیاس‌پذیری بالاتر و مقاومت بهتر در مقایسه با طرح‌های مطرح در سال‌های اخیر همچون SBIBD ، TD ، Trade-KP، *UKP، *RD و 2-D μ-PBIBD رسیده، همچنین هرچند که مدل پیشنهادی دارای اتصال‌پذیری کامل نیست، اما از اتصال‌پذیری بالایی نیز برخوردار می‌باشد.

## ۷- مراجع

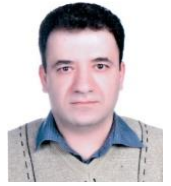

**اکبر مرشد اسکی،** تحصیلات خود را در دکتری مهندسی کامپیوتر گرایش نرم افزار به انجام رسانده است و هم اکنون استادیار در دانشگاه آزاد ورامین — پیشوا می‌باشد. تحقیقات مورد علاقه ایشان در زمینه‌های امنیت شبکه IoT و داده کاوی است. آدرس پست الکترونیکی ایشان عبارت است: morshed486@gmail.com

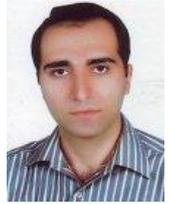

**سید حمید حاج سید جوادی،** تحصیلات خود را در دکتری ریاضی و علوم کامپیوتر به انجام رسانده است و هم اکنون استاد تمام در دانشگاه شاهد می‌باشد. تحقیقات مورد علاقه ایشان در زمینه‌های امنیت شبکه، الگوریتم و جبر کامپیوتری است. آدرس پست الکترونیکی ایشان عبارت است: h.s.javadi@shahed.ac.ir

# A novel key pre-distribution scheme based on µ-PBIBD combinatorial design in the resource-constrained IoT network

**Akbar Morshed Aski[1], Hamid Haj Seyyed Javadi[2]**

[1] Department of Computer Engineering, Varamin – Pishva Branch, Islamic Azad University, Varamin, Tehran, Iran
[2] Department of Mathematics and Computer Science, Shahed University, Tehran, Iran

## Abstract

In a resource-constrained IoT network, end nodes like WSN, RFID, and embedded systems are used which have memory, processing, and energy limitations. One of the key distribution solutions in these types of networks is to use the key pre-distribution scheme, which accomplishes the key distribution operation offline before the resource-constrained devices deployment in the environment. Also, in order to reduce the shared key discovery computing and communication overhead, the use of combinatorial design in key pre-distribution has been proposed as a solution in recent years. In this study, a µ-PBIBD combinatorial design is introduced and constructed and the mapping of such design as a key pre-distribution scheme in the resource-constrained IoT network is explained. Through using such key pre-distribution scheme, more keys are obtained for communication between two devices in the IoT network. This means that there will be a maximum of q + 2 keys between the two devices in the network, where q is the prime power, that is, instead of having a common key for a direct secure connection, the two devices can have q + 2 common keys in their key chain. Accordingly, we would increase the resilience of the key pre-distribution scheme compared to the SBIBD, TD, Trade-KP, UKP *, RD * and 2-D µ-PBIBD designs.

**Keywords**: resource-constrained IoT network; combinatorial design; µ-PBIBD; resilience.